\newif\ifcheckpagelimits
\checkpagelimitstrue
\checkpagelimitsfalse

\ifcheckpagelimits
 \documentclass[nofootinbib,pra,aps,twocolumn,showpacs,showkeys,%
 amsmath,amssymb,superscriptaddress,final,reprint,floatfix,longbibliography]{revtex4-1}
 \newcommand{\todo}[1]{}
\else
 \documentclass[pra,aps,twocolumn,showpacs,showkeys,%
 amsmath,amssymb,superscriptaddress,final,reprint,floatfix,longbibliography]{revtex4-1}
 \newcommand{\todo}[1]{{\pdfmargincomment[icon=Note,color=pink]{#1}}}
\fi

\usepackage{lineno}
  \usepackage{mathptmx}
\usepackage{subfigure}
\usepackage{dcolumn}
\usepackage{amsmath,amssymb}
\usepackage{bm}
\usepackage{color}
\usepackage{overpic}
\usepackage{latexsym}
\usepackage{epstopdf}
\usepackage{color}
\usepackage[english]{babel}
\usepackage{latexsym}
\usepackage{stmaryrd}

\usepackage{braket}

\definecolor{mygrey}{gray}{0.35}
\definecolor{myblue}{rgb}{0.2,0.2,0.8}
\definecolor{myzard}{cmyk}{0,0,0.05,0}
\definecolor{mywhite}{rgb}{1,1,1}
\definecolor{myred}{rgb}{1,0.,0.3}

\usepackage[colorlinks=true,citecolor=myblue,linkcolor=myred]{hyperref}

 \def\ee{\mathord{\rm e}}
 
 \def\ii{\mathord{\rm i}}

\renewcommand{\ii}{{\rm i}}
\renewcommand{\ee}{{\rm e}}
\def\beq{\begin{equation}}
\def\eeq{\end{equation}}

\begin{document}

\title{Quantum   Transport of Energy  in Controlled Synthetic  Quantum Magnets}

\author{Alejandro Bermudez}
\affiliation{Instituto de F\'{\i}sica Fundamental, IFF-CSIC, Calle Serrano 113b, Madrid E-28006, Spain}

\author{Tobias Schaetz}
\affiliation{Physikalisches Institut, Albert-Ludwigs-Universit\"{a}t,
Hermann-Herder-Stra\ss e 3, 79104 Freiburg, Germany}
\affiliation{Freiburg Institute for Advanced Studies (FRIAS),
Albertstrasse 19, 79104 Freiburg, Germany}

\begin{abstract}
We introduce  a scheme that exploits laser cooling and phonon-mediated spin-spin interactions in crystals of trapped atomic ions  to explore the  transport of energy through a  quantum magnet. We show how  to implement an effective transport window  to control the flow of energy through the magnet even in the absence of fermionic statistics for the carriers. This is achieved by shaping the  density of states of the effective thermal reservoirs that arise from the interaction with the external bath of the modes of the electromagnetic field, and  can be experimentally controlled by tuning the   laser frequencies and intensities appropriately. The interplay of this transport window with the  spin-spin interactions is exploited to build an analogue of the Coulomb-blockade effect in nano-scale electronic devices, and opens new possibilities to study quantum effects in energy transport.

\end{abstract}

%\pacs{TBD}
\ifcheckpagelimits\else
\maketitle
\fi
\setcounter{tocdepth}{2}
\begingroup
\hypersetup{linkcolor=black}
\tableofcontents
\endgroup

 \section{\bf Introduction} 
 
Richard Feynman's visionary character has served as a source of inspiration for physicists  of  different fields. In his  famous talk  ``There's plenty of room at the bottom''~\cite{feynman_porb}, he identified a wide and diverse number of technical applications based on manipulating and controlling devices at very small scales, ultimately, {\it atom by atom}. Independently of their technological prospects, these small devices   have the potential of displaying  utterly different phenomena from their large-scale counterparts, the understanding of which lies at the forefront of current fundamental research. This situation is clearly exemplified by experiments in nanoscale electronic devices~\cite{roukes_porb}, such as  the quantisation of electrical conductance~\cite{electric_transport_quantization}, and  the control of currents at the level of  single electrons~\cite{coulomb_blockade}. These   non-equilibrium processes  are ruled by  the laws of quantum mechanics, and can be considered as paradigms in  the field of {\it quantum transport}~\cite{nazarov_blanter}.

Controlling energy (heat) transport in small-scale devices, and understanding its similarities and differences with respect to semiclassical Boltzmann-type theories in macroscopic materials~\cite{rmp_heat_transport_phonons}, is also a  topic of fundamental and technological importance. Especially in the context of the shrinking  computer hardware, where  understanding and mitigating the mechanisms for the increasing energy dissipation is of the utmost importance. Even if the experimental study of energy transport in nano-devices is more challenging than its electronic counterpart~\cite{roukes_porb, rmp_heat_transport}, landmarks of quantum transport have also been achieved, such as the measurement of the  quantum of thermal conductance~\cite{thermal_transport_quantization}. However, to the best of our knowledge, controlling heat transport at the level of single energy quanta,  in analogy to the above electron Coulomb-blockade experiments~\cite{coulomb_blockade}, has not been considered so far. The purpose of this work is to study the occurrence of such phenomenon by considering the energy transport through an insulating quantum magnet, and to draw interesting  analogies with  the electronic transport in metallic nano-scale devices.

Although energy flow is generally dominated by phonon transport~\cite{rmp_heat_transport_phonons},  other carriers can also have large contributions, such as  electrons in   nano-devices as  a consequence of the large electron densities rather than a reduced temperature~\cite{rmp_heat_transport}. In insulating magnetic materials, the contribution of spin excitations (i.e. magnons) to the energy transport has been identified by measuring thermal conductances under different temperatures and magnetic fields~\cite{heat_tarnsport_spin_waves,review_magnon_heat_transport}. Low-dimensional cuprate compounds provide a  well-suited testbed to study spin-mediated heat transport~\cite{spinon_heat_transport_1d}, as the anisotropy  allows to distinguish the contributions to the thermal conductivity of the spin excitations (i.e. spinons) from that of the phonons~\cite{review_thermal_transport_spins_exp}. Moreover, these experiments link to a large body of theoretical studies that explore the relation of anomalous  transport with the integrability  of paradigmatic Heisenberg-type spin models~\cite{review_hermal_transport_spins_theory}. Nonetheless, these studies focus on  energy transport through macroscopically large magnets, whereas we are interested in  microscopic  devices which can be controlled at  a much smaller scale, ultimately, {\it spin by spin}. Just as the spin degrees of freedom  in single-molecule magnets can be exploited to control electronic currents~\cite{molecular_spintronics} for molecular electronics~\cite{molecular_elec_book}, we aim at controlling and studying the quantum transport of energy at the single quantum level by exploiting the spin degrees of freedom of an insulating quantum magnet.

To analyse a situation of experimental relevance, we shall rely on another of Feynman's visions. In his lecture ``Simulating physics with computers''~\cite{feynman_qs}, he put forth the possibility of mimicking the behaviour of a complex quantum model by a different, exquisitely controlled, quantum device. This idea, which has the potential of solving long-standing open questions in the field of quantum many-body physics~\cite{qs_goals}, has already found several applications within the realm of atomic, molecular and optical physics~\cite{QS_cold_atoms,QS_trapped_ions}. These experiments can be considered as performed in some sort of artificial  quantum matter, synthesised to behave according to the models that are supposed to capture the essence of the complex phenomena observed in the   condensed-matter systems.
 In particular, Coulomb crystals of trapped atomic ions~\cite{wineland_review}  offer a versatile playground to implement a range of synthetic one-dimensional quantum magnets~\cite{porras_spin_models_ions} that can be designed spin by spin~\cite{Ising_exp_ions}. We will introduce a particular scheme  to study energy transport in such synthetic quantum magnets, and  discuss the appearance of genuine quantum effects through analogues of  the well-known electron Coulomb-blockade physics. Although we shall focus on the one-dimensional case, the connectivity of the effective spin models can be changed by considering the two-dimensional magnets realised in Penning traps~\cite{penning_exp}, the generalisation of the  separate-well quantum magnets~\cite{qs_separate_wells} to two-dimensional surface traps~\cite{qs_surface_traps}, or the flexibility of digital approaches to quantum simulations in linear traps~\cite{digital_qs}.
  
This article is organised as follows. In Sec.~\ref{energy_transport}, we describe a general microscopic model to focus on energy transport through magnetic materials. In Sec.~\ref{sec:trapped_ions_energy_tranport}, we discuss in detail how all the required ingredients of such a  microscopic model can be implemented using Coulomb crystals of trapped atomic ions, and exploiting available tools that have been developed for high-precision metrology and quantum information processing. We also derive a master equation for quantum transport that describes the energy flow through a synthetic trapped-ion magnet, which is used to explore the energy-transport analogue of Coulomb blockade and single-electron transport in Sec.~\ref{sec:blockade}. In Sec.~\ref{sec:measurement}, we discuss a possible strategy to detect the relevant observables and to control and probe the energy transport in the trapped-ion setup, and we present our conclusions and outlook in Sec.~\ref{sec:outlook}.

\section{\bf Energy transport in quantum  magnets}
\label{energy_transport}

A typical experiment to study energy transport in magnetic materials~\cite{review_thermal_transport_spins_exp} in general relies on introducing a heat source that acts as a reservoir of lattice vibrations, whose energy is then transferred onto the magnetic degrees of freedom by the inherent crystalline spin-phonon coupling (e.g. scattering of phonons by paramagnetic ions). In analogy to the theory of electron transport, we consider a pair of macroscopically-large collections of harmonic oscillators  represented by the phonon modes of the crystal that act as  source and drain reservoirs. Therefore, the reservoirs are described by the Hamiltonian
\beq
\label{reservoirs}
H_{\rm p}=H_{\rm p,S}+H_{\rm p,D}=\sum_{n}\omega_{n,\rm{S}}a^{\dagger}_{n,{\rm S}}a^{\phantom{\dagger}}_{n,{\rm S}}+\sum_{n}\omega_{n,\rm{D}}a^{\dagger}_{n,{\rm D}}a^{\phantom{\dagger}}_{n,{\rm D}},
\eeq
where $a^{{\dagger}}_{n,r},a^{\phantom{\dagger}}_{n,r}$ are the  creation annihilation operators  of phonons with frequency $\omega_{n,r}$ in the $r\in\{\rm S,D\}$ source/drain reservoir labelled by a quantum number $n$ (e.g.  momentum in translationally-invariant systems), and we set $\hbar=1$. The source and drain oscillators shall induce a non-equilibrium energy current through the magnet under biased thermal conditions $\rho_{\rm p}^{\rm eq}\propto\ee^{-H_{\rm p,S}/k_{\rm B}T_{\rm S}}\otimes\ee^{-H_{\rm p,D}/k_{\rm B}T_{\rm D}}$, where $T_{\rm S}>T_{\rm D}$, and $k_{\rm B}$ is Boltzmann's constant. Moreover, to qualify as reservoirs, their state should not be modified by the coupling to the magnetic system $\rho_{\rm p}(t)={\rm tr}_{\rm m}\{\rho(t)\}=\rho_{\rm p}(0)$ $\forall t$, which is typically justified by considering the macroscopically-large number of degrees of freedom of the reservoirs~\eqref{reservoirs} in comparison to those of a smaller quantum system they are connected to.

The insulating magnetic system will be described by some microscopic spin-chain model 
\beq
\label{spin_chain_model}
H_{\rm m}=\sum_ih_i+\sum_{i,j}h_{i,j},
\eeq
which represents the spin-1/2 particles  by Pauli matrices  $\boldsymbol{\sigma}_i=(\sigma_i^x,\sigma_i^y,\sigma_i^z)$, subjected to local terms (e.g. a transverse magnetic field $h_i=-h\sigma_i^x$ ), and pairwise interactions (e.g. Heisenberg  $h_{i,j}=J_{ij}\boldsymbol{\sigma}_i\cdot \boldsymbol{\sigma}_j$, or Ising  $h_{i,j}=J_{ij}\sigma_i^z {\sigma}^z_j$ couplings).

%%%%%%%%%%%%%%%%%%%%%%%%%%%%%%%%%%%%%%%%%%%%%%%%%%%%%%%%%%%%
\begin{figure}
\centering
\includegraphics[width=0.9\columnwidth]{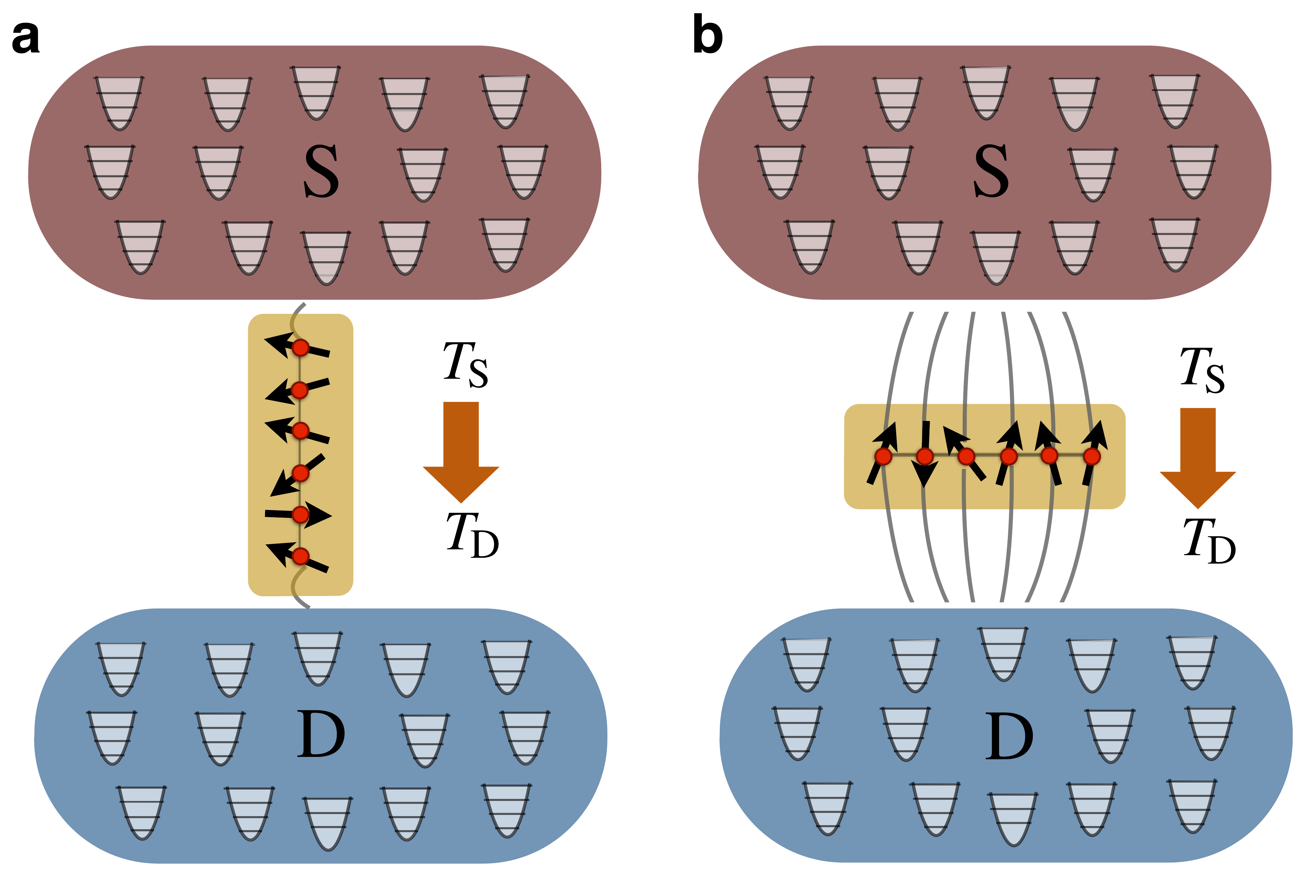}
\caption{ {\bf Energy transport in quantum magnets:} Biased source/drain reservoirs represented by a macroscopic number of harmonic oscillators in a thermal state  with temperatures $T_{\rm S}>T_{\rm D}$, such that an energy current through a quantum magnet is established. The quantum magnet is depicted as a chain of interacting spins that can exchange energy with the reservoirs, such that transport occurs longitudinally along the chain {\bf (a)}, or transversely across it {\bf (b)}.  }
\label{fig_scheme}
\end{figure}
%%%%%%%%%%%%%%%%%%%%%%%%%%%%%%%%%%%%%%%%%%%%%%%%%%%%%%%%%%%%

As stated above, it is crucial that some microscopic mechanism provides a spin-phonon coupling that allows energy to be transferred between the reservoirs and the magnetic system.  Although there might be spin-Peierls-type couplings responsible for the energy exchange~\cite{review_thermal_transport_spins_exp}, which could also be synthesised  in the trapped-ion magnet~\cite{spin_peierls_ions},  we shall rely on a generic spin-phonon scattering mechanism that can be described again as a pairwise spin-phonon coupling
\beq
\label{energy_exchange}
H_{\rm mp}=\sum_{n,i}g_{n,i,{\rm{S}}}S^{\phantom{\dagger}}_ia^{{\dagger}}_{n,{\rm S}}+\sum_{n,i}g_{n,i,{\rm{D}}}S^{\phantom{\dagger}}_ia^{{\dagger}}_{n,{\rm D}}+{\rm H.c.},
\eeq
where we have introduced the source-drain spin-phonon couplings $g_{n,i,r}$, and a general spin operator $S^{\phantom{\dagger}}_i$ that should induce a transition between two eigenstates $\ket{\epsilon_{\ell}}\to\ket{\epsilon_{\ell'}}$ of the above spin-chain model~\eqref{spin_chain_model}, such that energy can be exchanged between the vibrational and magnetic degrees of freedom. Within the so-called orthodox theory of quantum electronic transport~\cite{orthodox_theory,single_electron_review}, the corresponding  transition rates, $\Gamma(\ell,\ell')$,  are obtained applying Fermi's golden rule. One can thus calculate such rates for our current problem, and use them later to describe the energy transport across the quantum magnet.

In the following, we shall use this section as a guiding principle, and discuss the particular trapped-ion realisation of the above ingredients, namely {\it (a)} the source-drain biased reservoirs, {\it (b)} the synthetic quantum magnet,  {\it (c)} the engineered spin-phonon coupling for the energy exchange mechanism, and {\it (d)} the analogue of the Fermi golden rule transition rates used to study the transport. According to the dependence of the spin-phonon couplings $g_{n,i,r}$ on the spin site index $i$, we can model either the longitudinal energy transport  [cf. Fig.~\ref{fig_scheme}{\bf (a)}], or  transverse energy transport  [cf. Fig.~\ref{fig_scheme}{\bf (b)}].

\section{\bf Coulomb crystals for energy transport}
\label{sec:trapped_ions_energy_tranport}

Before embarking on the above goal, let us start  by reviewing the studies on energy transport through trapped-ion Coulomb crystals that have already appeared in the literature.  In the case of lattice vibrations, one possibility is to consider local quenches, namely pump-probe experiments where an initial excited state with an inhomogeneous energy density is prepared by some local perturbation, and its evolution under the microscopic vibrational Hamiltonian is  probed at different instants of time~\cite{prop_vib_excitations_th_ions,prop_vib_excitations_exp_ions}. Pump-probe experiments   have also been performed for the synthetic quantum magnets~\cite{global_quench_spins_trapped_ions, dynamics_spins_trapped_ions}. Interestingly, for the local quenches~\cite{dynamics_spins_trapped_ions},  by measuring the spread of certain variances for a particular local perturbation~\cite{real_time_conductivities}, these experiments  could address the effect of  long-range terms, or  additional perturbations, on the anomalous energy transport predicted for the integrable nearest-neighbour XY model~\cite{review_hermal_transport_spins_theory}. However,  finite-size effects associated to the small number of spins may obscure the results.

The other possibility relies on the more standard transport scenario, where a pair of temperature-biased  reservoirs is connected to the system~\eqref{reservoirs}. Unfortunately, the ion Coulomb crystals typically considered in this context are rather small, and the number of phonon modes differs markedly from the required macroscopic number of degrees of freedom of a reservoir~\eqref{reservoirs}. On the other hand,  more fundamentally, the key property of an energy reservoir is that it should be capable of supplying/absorbing  arbitrary amounts of energy  into/from the system without being modified. Provided that the vibrational modes of the ions display this property, which can be controlled by means of laser cooling and heating,  one may consider them as effective thermal reservoirs despite their finite number. In some sense, the laser cooling couples the vibrational modes to the infinite number of photonic modes in the electromagnetic bath, such that one obtains an effective reservoir. So far, this has only been considered for phonon-mediated energy transport in trapped-ion crystals, where the heat reservoirs correspond  either to individually-addressed laser-cooled ions~\cite{prop_vib_excitations_th_ions,ion_chain_two_reservoirs_thermalization,heat_transport_zz_ions,heat_transport_hubard_ions}, or to ions of a different species  in sympathetically-cooled crystals~\cite{heat_transport_sympathetic_toolbox_ions}.  

As analysed in~\cite{heat_transport_sympathetic_toolbox_ions} (see the detailed \href{http://journals.aps.org/prl/supplemental/10.1103/PhysRevLett.111.040601}{Supplemental Material} of that paper), there are some rather stringent conditions on the cooling rates that must be fulfilled for the laser-cooled ions to resemble a canonical transport reservoir. In particular, the cooling rates must be much larger than the vibrational couplings between distant ions. As a consequence,   standard Doppler cooling by a travelling wave in a  linear Paul trap does not suffice, and other cooling schemes must be adopted, such as standing-wave cooling~\cite{laser_coolig_sw} or electromagnetic-induced-transparency (EIT) cooling~\cite{eit_cooling}. Although  EIT cooling has been demonstrated~\cite{eit_cooling_exp}, both schemes add on to the complexity of the  transport setup, and  it would be desirable to devise new  protocols where travelling-wave Doppler cooling suffices.

According to all the above discussions, we identify a two-fold interest in focusing on the energy transport through a trapped-ion synthetic magnet, rather than via lattice vibrations. As shown below,  the energy transport setup shall only require travelling-wave Doppler cooling. Besides, the intrinsic non-linearities associated to the synthetic magnet shall yield a heat analog of Coulomb-blockade physics, paving the way to access a so far neglected, yet relevant and new, quantum effect in energy transport.

Once this has been discussed, we can embark upon the description of the trapped-ion realisations of {\it (a)} the source-drain biased reservoirs, {\it (b)} the synthetic quantum magnet,  {\it (c)} the engineered spin-phonon coupling for the energy exchange, and {\it (d)} the analogue of the Fermi golden rule transition rates used to study the energy transport.

\subsection{Collective  transport reservoirs}
\label{sec:reservoirs}

We  consider a mixed Coulomb crystal of N atomic ions of two different species/isotopes (e.g.  $^{25}{\rm Mg}^+$ and $^{26}{\rm Mg}^+$), confined in a linear Paul trap with  frequencies $\omega_x\neq\omega_y\gg\omega_z$. The collective lattice vibrations of this crystal can be described in terms of three phonon branches 
\beq
H_{\rm p}=\sum_{n,\alpha}\omega_{n,\alpha}a_{n,\alpha}^{\dagger}a_{n,\alpha}^{\phantom{\dagger}},
\eeq
where we have introduced the  vibrational frequencies $\omega_{n,\alpha}$  for each normal  mode $n\in\{1,\dots,N\}$ in each branch $\alpha\in\{x,y,z\}$, and the creation-annihilation operators $a_{n,\alpha}^{\dagger}, a_{n,\alpha}^{\phantom{\dagger}}$ of phonons for each of those frequencies. The ions are subjected  to  a  laser beam  tuned  close to the resonance of a dipole-allowed transition of one of the atomic  species, which shall be referred to as the ionic coolant (e.g. $^{26}{\rm Mg}^+$), responsible for sympathetic Doppler cooling to feature the characteristics of a energy-transport reservoir. However, we consider a regime opposite to our previous work~\cite{heat_transport_sympathetic_toolbox_ions}, by focusing on  individual laser-cooling rates~\cite{laser_coolig_sw}  much smaller than the vibrational couplings between both species. In this limit, laser cooling pumps the collective vibrational modes, and not the local vibrations~\cite{heat_transport_sympathetic_toolbox_ions},  into a thermal steady state.  This eases the requirements substantially  on the cooling schemes discussed above, and allows us to consider the established travelling-wave cooling. 

If the travelling-wave beam is directed along the axis of the trap, it induces a damping of the longitudinal phonons  described by a dissipation  Lindblad-type~\cite{oqs_book} super-operator, namely 
\beq
\label{cooling_dissipator}
\mathcal{D}_{\rm p}(\rho_{\rm p})=\sum_{n}\sum_{s=+,-}\left({L_{n,s}\rho_{\rm p} L^{\dagger}_{n,s}-L_{n,s}^{\dagger}L_{n,s}\rho_{\rm p}}\right)+{\rm H.c.}, 
\eeq
with the following jump operators 
\beq
\label{jump_op}
L_{n,+}=\sqrt{\Gamma_{n, +}}a_{n,z}^{\dagger},\hspace{2ex} L_{n,-}=\sqrt{\Gamma_{n, -}}a_{n,z}^{\phantom{\dagger}},
\eeq
where the heating (cooling) rates $\Gamma_{n, +} (\Gamma_{n, -})$ can be obtained from the  laser-cooling rates of a single ion $\Gamma_{\pm}(\omega_{\rm t})$~\cite{laser_coolig_sw} with a trap frequency corresponding to the normal mode frequency $\omega_{\rm t}=\omega_{n,z}$, after considering the normal-mode displacements $\mathcal{M}_{i,n}^z$ at the positions of the ionic coolants, 
\beq
\label{heating_cooling}
\Gamma_{n, \pm}=\sum_{i}(\mathcal{M}_{i,n}^z)^2\frac{\omega_z}{\omega_{n,z}}\Gamma_{\pm}(\omega_{n,z}).
\eeq

Since the longitudinal modes are well separated in frequencies, we can focus on a couple of modes, which will play the role of the source and drain thermal reservoirs $n_{\rm S}, n_{\rm D}\in\{1,\cdots, N\}$. When the laser beam is red detuned from the  dipole-allowed transition, $\Gamma_{n, +} <\Gamma_{n, -}$,  the normal modes are Doppler cooled to the desired  steady state, 
\beq
\label{cooling_steady_state}
\rho_{\rm p}^{\rm eq}=\frac{\ee^{-H_{\rm p,S}/k_{\rm B}T_{\rm S}}}{{\rm tr}\{\ee^{-H_{\rm p,S}/k_{\rm B}T_{\rm S}}\}}\otimes\frac{\ee^{-H_{\rm p,D}/k_{\rm B}T_{\rm D}}}{{\rm tr}\{\ee^{-H_{\rm p,D}/k_{\rm B}T_{\rm D}}\}},
\eeq where we have introduced the temperatures 
\beq
\label{temperatures}
T_{r}=\frac{\omega_{r}}{\log \{(\bar{n}_{r}+1)/\bar{n}_{r}\}},\hspace{2ex}\bar{n}_{r}=\frac{\Gamma_{n, +}}{\Gamma_{n, -}-\Gamma_{n, +}},
\eeq
and defined $\omega_r:=\omega_{n_{r},z}$ to simplify the notation. The equilibrium temperatures of the reservoirs, as well as the temperature bias $\delta T=T_{\rm S}-T_{\rm D}$, can be controlled to some extent by simply modifying the  detuning of the cooling laser, as detailed below in Fig.~\ref{fig_Ts}. Let us note that the range of the temperature bias could be extended by using two normal modes along different axes as the source/drain reservoirs. 

As far as the sympathetic Doppler cooling remains switched on  during the transport experiment, and the spin-phonon exchange mechanism is weaker than the overall cooling rate, the vibrational state of the two modes remains frozen in the desired thermal state $\rho_{\rm p}(t)={\rm tr}_{\rm m}\{\rho(t)\}=\rho_{\rm p}^{\rm eq}$ $\forall t$, and the normal modes can be considered as an effective biased reservoir for  transport. Here, the energy supply of the source reservoir comes from the laser driving, whereas the absorption of the excess energy in the drain reservoir is stored in the electromagnetic bath through the dipole-allowed transition.

\subsection{Synthetic quantum magnet}

Once we have discussed the scheme to synthesise  the  biased reservoirs by laser-cooled collective modes, let us describe how an interacting spin chain~\eqref{spin_chain_model} can be implemented using an ion crystal~\cite{porras_spin_models_ions}. We encode the spins  $\ket{\uparrow},\ket{\downarrow}$ in two long-lived electronic states of the remaining species with  energy difference $\omega_0$ (e.g. in two hyperfine states $\ket{F,M}$ of the $^{25}{\rm Mg}^+$ groundstate manifold, $\ket{\uparrow}=\ket{2,2},\ket{\downarrow}=\ket{3,3}$). Therefore, the sums in Eq.~\eqref{spin_chain_model}  must be restricted to the sites of the crystal  where electronic degrees of freedom remain unaffected, namely to the sites of the non actively-cooled atomic ions.

To induce the spin-spin interaction, we use  state-dependent dipole forces that push the ions along  the radial $x,y$ axes, and provide a spin-phonon coupling with the vibrational modes that are not affected by the continuously laser cooling of the longitudinal branch. While near-resonant forces are already established to implement two-qubit gates for quantum information processing~\cite{molmer_sorensen, didi_gate, bermudez_gate}, far-detuned forces~\eqref{spin_chain_model} lead to interacting spin models  where the radial phonons  act as  carriers of the spin-spin interactions that only get excited virtually, and can be adiabatically eliminated from the dynamics. The local terms in Eq.~\eqref{spin_chain_model} correspond to ac-Stark shifts or carrier transitions~\cite{wineland_review}. 

The paradigmatic synthetic  spin chain~\eqref{spin_chain_model}, already implemented in a variety of experiments~\cite{Ising_exp_ions}, is the quantum Ising model
\beq
\label{Ising}
h_i=-h\sigma_i^x,\hspace{2ex}h_{i,j}=J_{ij}\sigma_i^z\sigma_j^z,
\eeq
which only requires a carrier for the transverse field, and a state-dependent force for the Ising interactions.
Note that the relative magnitude of both terms, and the range of the antiferromagnetic interactions, can be experimentally controlled.

Using a couple of state-dependent forces, each along a different axes, and exploiting the different trap frequencies $\omega_x\neq\omega_y$~\cite{porras_spin_models_ions}, it is also possible to realise the anisotropic XY model in a tunable transverse field
\beq
\label{XY}
h_i=-h\sigma_i^z,\hspace{2ex}h_{i,j}=J^x_{ij}\sigma_i^x\sigma_j^x+J^y_{ij}\sigma_i^y\sigma_j^y,
\eeq
where the transverse field now requires an ac-Stark shift, and all the relative magnitudes can be engineered experimentally. A particular limit, the isotropic XY model in a strong transverse field $h\gg J_{i,j}^x=J_{ij}^y$, can also be obtained by exploiting a single state-dependent dipole force under a strong transverse field, as demonstrated experimentally in~\cite{global_quench_spins_trapped_ions, dynamics_spins_trapped_ions}. To get access to the anisotropic regime with a single dipole force, one needs to  rapidly modulate the transverse field periodically in time, as proposed in~\cite{duality}.

If we combine these ideas with another state-dependent force along the remaining axes in the same basis as the strong local term, it is possible to engineer the XXZ model
\beq
\label{XXZ}
h_i=-h\sigma_i^z,\hspace{2ex}h_{i,j}=J^\perp_{ij}(\sigma_i^x\sigma_j^x+\sigma_i^y\sigma_j^y)+J_{ij}^{||}\sigma_i^z\sigma_j^z,
\eeq
where the transverse field is stronger than the spin-spin couplings. The dynamics of such models in a strong transverse field~\cite{global_quench_spins_trapped_ions,dynamics_spins_trapped_ions}, or the transport transport phenomena they can give rise to, can be indeed highly non-trivial and interesting.

The last, and most involved possibility, would be consider three state-dependent dipole forces  along all of the axes~\cite{porras_spin_models_ions}, which would lead to a XYZ model
\beq
\label{XYZ}
h_i=-h\sigma_i^z,\hspace{2ex}h_{i,j}=J^x_{ij}\sigma_i^x\sigma_j^x+J^y_{ij}\sigma_i^y\sigma_j^y+J_{ij}^{z}\sigma_i^z\sigma_j^z,
\eeq
where the ratio of the transverse field and all the remaining couplings remains tunable. In order to use the laser-cooled longitudinal modes also as carriers of a spin-spin coupling, one must employ larger detunings, at the expense of obtaining weaker spin-spin couplings~\cite{dissipation_assisted}. To study the energy transport through any of these synthetic quantum magnets, we now need to discuss the energy transfer  between the spins and the laser-cooled phonons playing the role of reservoirs.

\subsection{Spin-phonon energy exchange}

\begin{figure}
\centering
\includegraphics[width=1\columnwidth]{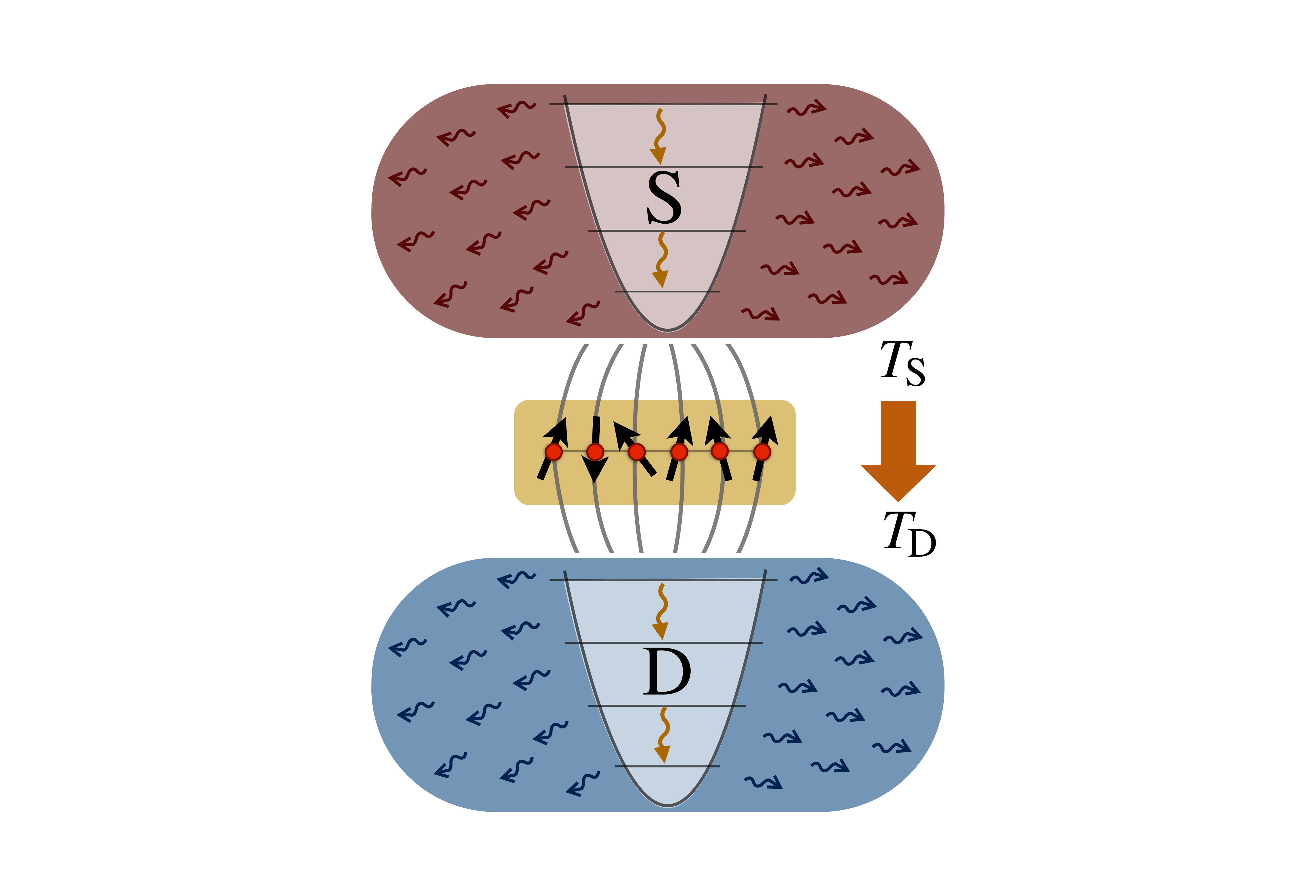}
\caption{ {\bf Energy transport in an ion quantum magnet:} The effective transport reservoirs correspond, in the trapped ion setup, to only a couple of normal modes that are laser cooled to different temperatures. The interacting spin model is composed of pseudo-spins corresponding to a couple of electronic levels, such that spin-spin interaction can be mediated by phonons. The energy exchange  between the synthetic magnet and the reservoirs occurs transversally to the effective spin chain, however, best mediated via axial laser excitations in the resolved-sideband regime.}
\label{fig_scheme_ions}
\end{figure}

Since the longitudinal vibrational modes of an ion chain are well-separated in frequency, we can consider  a pair of laser excitations (e.g. each provided by  two  Raman beams for $^{25}{\rm Mg}^+$),  with frequencies  $\omega_{{\rm L},{\rm S}}\approx\omega_0-\omega_{\rm S}$, $\omega_{{\rm L},{\rm D}}\approx\omega_0-\omega_{\rm D}$, tuned close to the red sideband~\cite{wineland_review} of two different normal modes   $n_{\rm S}, n_{\rm D}\in\{1,\cdots, N\}$. In the resolved-sideband limit, these terms are analogous to the required energy-exchange mechanism in Eq.~\eqref{energy_exchange}, namely
 \beq
\label{energy_exchange}
H_{\rm mp}(t)=\sum_{i}g_{i,{\rm{S}}}S^{\phantom{\dagger}}_ia^{\phantom{\dagger}}_{n_{\rm S},z}\ee^{-\ii\delta_{{\rm S}}t}+\sum_{i}g_{i,{\rm{D}}}S^{\phantom{\dagger}}_ia^{\phantom{\dagger}}_{n_{\rm D},z}\ee^{-\ii\delta_{{\rm D}}t}+{\rm H.c.},
\eeq
where $S_i=\sigma_i^+:=\ket{{\uparrow}_i}\bra{{\downarrow}_i}$. Here, the energy-exchange couplings $g_{i,r}$ can be obtained from the individual red-sideband couplings $g_i$~\cite{wineland_review}  by considering the normal-mode frequencies and displacements $g_{i,r}=g_i\mathcal{M}^z_{i,n_{r}}\sqrt{\omega_z/\omega_{n_{r},z}}$,  the laser detunings correspond to  
\beq
\label{laser_detunings}
\delta_{{r}}=\omega_{{\rm L},{\rm r}}-(\omega_0-\omega_{n_{r},z}),
\eeq and we are working in an interaction picture. Since each of the reservoirs is, in principle, coupled to all of the spins in the synthetic magnet, the trapped-ion transport scheme depicted in Fig.~\ref{fig_scheme_ions} resembles the transverse energy transport [cf. Fig.~\ref{fig_scheme}{\bf (b)}]. 

Once the trapped-ion energy exchange mechanism has been introduced, and the expression of the energy-exchange couplings $g_{i,r}$ explicitly given, we can revisit the crucial point raised in Sec.~\ref{sec:reservoirs}: to consider the two vibrational modes as reservoirs, their  state $\rho_{\rm p}(t)={\rm tr}_{\rm m}\{\rho(t)\}=\rho_{\rm p}^{\rm eq}$ $\forall t$  must remain unperturbed~\eqref{cooling_steady_state}, regardless of their coupling with the magnetic system. This sets a limitation on the possible strengths of the energy-exchange couplings with respect to the cooling rates
\beq
\label{cond_reservoir}
|g_{i,r}|\ll \kappa_{r}:=2{\rm Re}\{\Gamma_{n_{r},-}-\Gamma_{n_{r},+}\},
\eeq
which implies that the resulting damping rates $\kappa_{r}$ from the interplay of the cooling and heating processes in Eq.~\eqref{jump_op} must be much stronger. In this case, the state of the laser-cooled modes is effectively frozen, such that they play the role of the thermal reservoirs.
%%%%%%%%%%%%%%%%%%%%%%%%%%%%%%%%%%%%%%%%%%%%%%%%%%%%%%%%%%%%
%%%%%%%%%%%%%%%%%%%%%%%%%%%%%%%%%%%%%%%%%%%%%%%%%%%%%%%%%%%%

\subsection{Transition rates and transport window}

\begin{figure}
\centering
\includegraphics[width=0.95\columnwidth]{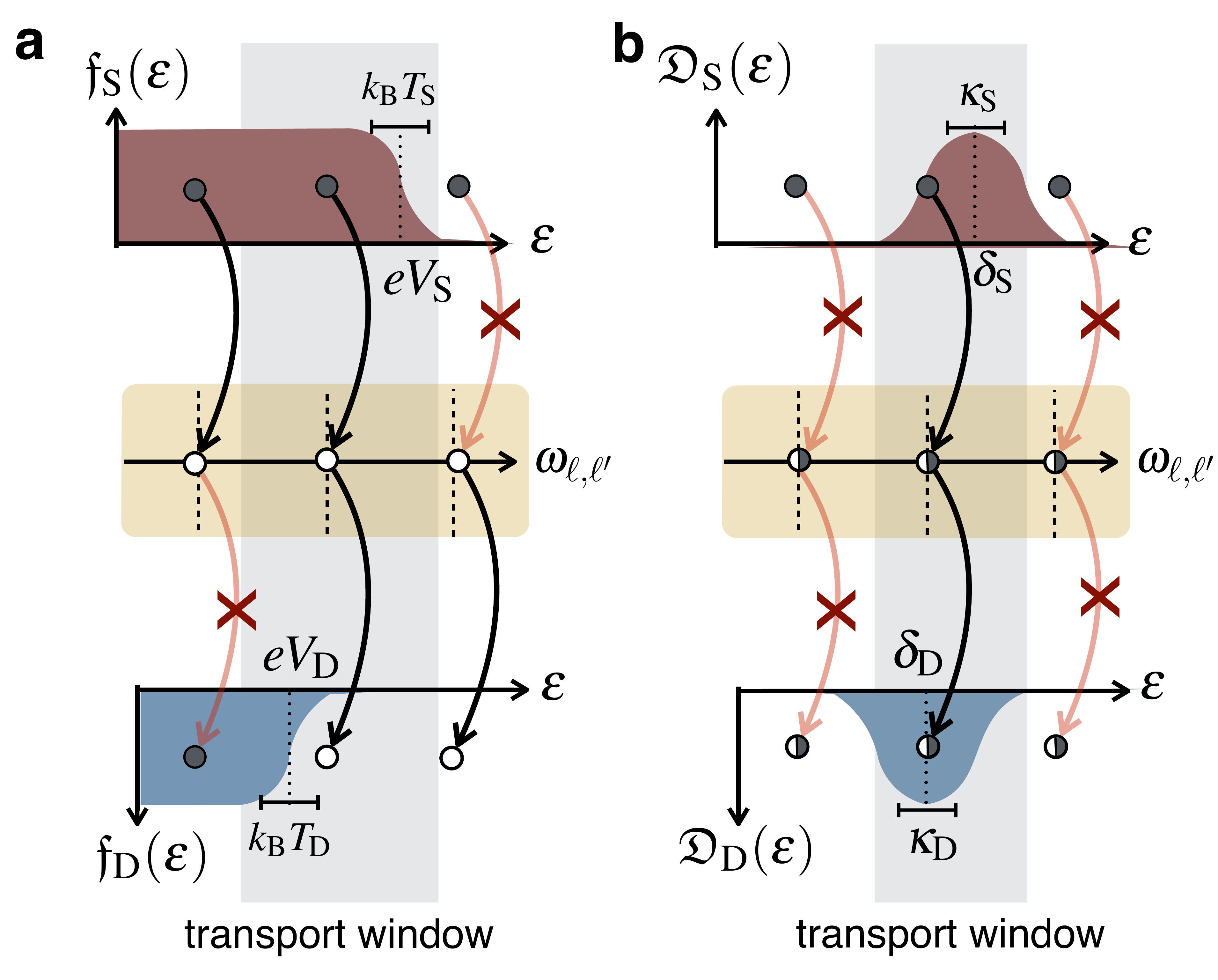}
\caption{ {\bf Transport window in electronic and energy currents:} Transport in nano-structures can be described by the  tunnelling of carriers from one reservoir onto the nano-device, and then onto the remaining reservoir. Such tunnelling events involve a transition between two energy levels  $\ket{\epsilon_{\ell'}}\to\ket{\epsilon_{\ell}}$ of the nano-device, with a transition frequency $\omega_{\ell,\ell'}=\epsilon_{\ell}-\epsilon_{\ell'}$. {\bf (a)} In electronic transport, current can flow through that particular channel if the transition frequency lies within the transport window, which is defined by the overlap of the  Fermi-Dirac distributions for source electrons $\frak{f}_{\rm S}(\epsilon)$, and drain vacancies $1-\frak{f}_{\rm D},(\epsilon)$ respectively, and is represented by a shadowed grey area in the figure. This window is  controlled by the bias voltages $V_{\rm S},V_{\rm D}$, and the leads temperatures $T_{\rm S}, T_{\rm D}$. {\bf (b)} In energy transport, a similar transport window can be defined by the overlap of the density of states of the source $\mathfrak{D}_{\rm S}(\epsilon)$ and drain $\mathfrak{D}_{\rm D}(\epsilon)$ reservoirs, which can be controlled by the parameters that control the maximum $\delta_{\rm S},\delta_{\rm D}$ and width $\kappa_{\rm S},\kappa_{\rm D}$ of the source-drain density of states.     }
\label{fig_transport_window}
\end{figure}

Having introduced  all the independently-controllable  ingredients of the  trapped-ion transport toolbox in the previous sections, we shall now describe how to calculate the analogue of the transition rates $\Gamma_{\rm i\to f}$ used in the conventional orthodox theory of quantum electronic transport~\cite{single_electron_review}. In this theory, such rates are calculated by means of Fermi's golden rule, and turn out to be proportional to the density of states and the Fermi-Dirac distribution of the metallic leads that act as  the  source/drain reservoirs. One typically assumes the wide-band limit, where the lead's density of states $\mathfrak{D}_{r}(\epsilon)$ is  featureless. On the other hand, the Fermi-Dirac distribution $\mathfrak{f}_r(\epsilon)$, which can be controlled by the external voltages  biasing the  leads, plays a fundamental role in determining the electric current in nano-scale devices: electrons can only propagate across the device provided that  the source reservoir features electrons of the required energy to tunnel into the device, while the drain reservoir has to provide  vacancies at the  energy of the electron trying to enter from the device. Therefore, Pauli exclusion principle defines the electronic transport window that is crucial for the theory of quantum transport in nanostructures [cf. Fig.~\ref{fig_transport_window}{\bf (a)}]. 
In contrast, the standard  setup of energy transport through  quantum magnets of Sec.~\ref{energy_transport},  involves a pair of temperature-biased bosonic reservoirs, such that one cannot rely  on the fermionic statistics to define a transport window. However, we can engineer the reservoir's density of states in order to obtain a similar transport window, despite the Bose-Einstein statistics of the phonons [cf. Fig.~\ref{fig_transport_window}{\bf (b)}], while achieving a full control of its characteristics. In the following, we  show how that is possible for the trapped-ion setups.

%%%%%%%%%%%%%%%%%%%%%%%%%%%%%%%%%%%%%%%%%%%%%%%%%%%%%%%%%%%%
%%%%%%%%%%%%%%%%%%%%%%%%%%%%%%%%%%%%%%%%%%%%%%%%%%%%%%%%%%%%

The calculation of the transition rates is more involved in this case, as we have a mixture of coherent and dissipative dynamics, which must be treated using the formalism of quantum master equations~\cite{oqs_book}.  The use of quantum master equations in the theory of electronic transport is an alternative to  more common approaches~\cite{molecular_elec_book}, such as the scattering formalism or non-equilibrium Green's functions, and is gaining more attention recently~\cite{timm_meq_transport}. In contrast to these other methods, and in addition to bringing up the possibility of describing  transient effects, quantum master equations have the advantage that strong-correlation effects within the device need not be treated perturbatively. In our case, the mixture of coherent and dissipative dynamics forbids applying  the general formalism~\cite{timm_meq_transport} directly. However, we show  that one can apply adiabatic elimination techniques in the theory of open quantum systems~\cite{ad_elimination} in this  context, and obtain a  quantum master equation for energy transport  through our synthetic magnet.

To proceed further, let us consider the formal diagonalization of the spin model~\eqref{spin_chain_model} corresponding to  any of the possible synthetic realisations~\eqref{Ising}-\eqref{XYZ}, namely
\beq
\label{diag_spin}
H_{\rm m}=\sum_{\ell}\epsilon_\ell\ket{\epsilon_\ell}\bra{\epsilon_\ell},
\eeq
where $\ket{\epsilon_\ell}$ are the magnetic eigenstates, and $\epsilon_\ell$ the associated energies. Working in an interaction picture that also includes the spin model, the energy-exchange term~\eqref{energy_exchange} becomes
\beq
\label{mp_spin_basis}
H_{\rm mp}(t)=\sum_{\ell,\ell'}\sum_{r}\tilde{g}_{\ell,\ell', r}J_{\ell,\ell'}a_{n_r,z}\ee^{\ii(\omega_{\ell,\ell'}-\delta_r)t},
\eeq
where we have introduced the transition operators, transition frequencies, and transition coupling strengths
\beq
J_{\ell,\ell'}=\ket{\epsilon_\ell}\bra{\epsilon_{\ell'}},\hspace{2ex}\omega_{\ell,\ell'}=\epsilon_\ell-\epsilon_{\ell'},\hspace{2ex}\tilde{g}_{\ell,\ell',r}=\sum_i{g}_{i,r}\bra{\epsilon_\ell}\sigma_i^+\ket{\epsilon_{\ell'}}.
\eeq
 The complete coherent and dissipative elements of the dynamics can then be expressed as a master equation $\dot{\rho}(t)=(\mathcal{L}_0+\mathcal{L}_1)\rho(t)$, where $\mathcal{L}_0(\rho)=\mathcal{D}_{\rm p}(\rho)$ is the laser-cooling super-operator~\eqref{cooling_dissipator}, and $\mathcal{L}_1(\rho)=-\ii[H_{\rm mp}(t),\rho]$ is the Liouville super-operator associated to Eq.~\eqref{mp_spin_basis}. Provided that the condition~\eqref{cond_reservoir} is fulfilled, we can assume that the effective reservoirs remain unchanged during the whole evolution, and thus adiabatically eliminate them from the dynamics by using projection-operator techniques~\cite{ad_elimination}.  To  lowest-order, the transport through the quantum magnet can be described by 
\beq
\label{ad_elim}
\dot{\rho}_{\rm m}(t)={\rm tr}_{\rm p}\left\{\int_0^\infty{\rm d}s\mathcal{P}\mathcal{L}_1(t)\ee^{\mathcal{L}_0s}\mathcal{L}_1(t-s)\mathcal{P}\rho(t)\right\}, 
\eeq
where we have introduced the projector $\mathcal{P}O=\rho_{\rm p}^{\rm eq}\otimes{\rm tr}_{\rm p}\{O\}$ onto the reservoirs steady state~\eqref{cooling_steady_state}, and   the reduced density matrix of the quantum magnet ${\rho}_{\rm m}(t)={\rm tr}_{\rm p}\{\rho(t)\}$.  

The resulting master equation for the magnetic degrees of freedom, which can be obtained by applying the quantum regression theorem to evaluate the lesser/greater single-particle Green's functions of the laser-cooled phonons, becomes
\beq
\label{transport_meq}
\dot{\rho}_{\rm m}(t)=-\ii\left[\sum_\ell(\Delta\epsilon_{\ell,{\rm S}}+\Delta\epsilon_{\ell,{\rm D}})\ket{\epsilon_\ell}\bra{\epsilon_\ell},{\rho}_{\rm m}(t)\right]+\mathcal{D}_{\rm m}(\rho_{\rm m}(t)),
\eeq
where we have introduced Lamb-type shifts of the quatum magnet energy levels~\eqref{diag_spin} caused by their coupling to the source/drain $r\in\{\rm S,D\}$ reservoirs 
\beq
\Delta\epsilon_{\ell,r}=-\sum_{\ell'}\frac{|\tilde{g}_{\ell,\ell',r}|^2(\delta_{r}-\omega_{\ell,\ell'})}{(\delta_{r}-\omega_{\ell,\ell'})^2+(\kappa_r/2)^2}.
\eeq
More relevant to the problem of quantum transport is the dissipative Lindblad-type super-operator
\beq
\label{transport_dissipator}
\mathcal{D}_{\rm m}(\rho_{\rm m})=\frac{1}{2}\sum_{\ell,\ell'}\sum_{s=+,-}\left({\tilde{L}_{\ell,\ell',s}^{\phantom{\dagger}}}\rho \tilde{L}_{\ell,\ell',s}^{\dagger}-\tilde{L}_{\ell,\ell',s}^{\dagger}{\tilde{L}_{\ell,\ell',s}^{\phantom{\dagger}}}\rho_{\rm m}\right)+{\rm H.c.},
\eeq
where we have introduced the jump operators
\beq
\begin{split}
\tilde{L}_{\ell,\ell',+}=\sqrt{\Gamma_{\rm SM}(\ell,\ell')+\Gamma_{\rm DM}(\ell,\ell')}J_{\ell',\ell}^{\dagger},\\
\tilde{L}_{\ell,\ell',-}=\sqrt{\Gamma_{\rm MS}(\ell,\ell')+\Gamma_{\rm MD}(\ell,\ell')}J_{\ell,\ell'}^{\phantom{\dagger}},
\end{split}
\eeq
which describe the quantum jumps $\ket{\epsilon_{\ell'}}\to\ket{\epsilon_{\ell}}$ by the transfer of an energy quantum from the reservoirs onto the magnet $\Gamma_{{\rm M}r}(\ell,\ell')$, or  from the magnet onto the reservoirs $\Gamma_{r{\rm M}}(\ell,\ell')$. The transition rates can be expressed as follows
\beq
\label{rates}
\begin{split}
\Gamma_{{\rm M}r}(\ell,\ell')&=2\pi|\tilde{g}_{\ell,\ell',r}|^2\mathfrak{D}_r(\omega_{\ell,\ell'})\frak{n}_r(\omega_r),\\
\Gamma_{r{\rm M}}(\ell,\ell')&=2\pi|\tilde{g}_{\ell',\ell,r}|^2\mathfrak{D}_r(\omega_{\ell',\ell})(1+\frak{n}_r(\omega_r)),
\end{split}
\eeq
which depend on the Bose-Einstein distribution of the thermal reservoirs $\frak{n}_r(\epsilon)=1/(\ee^{\epsilon/{k_{\rm B}T_r}}-1)$, and also on a Lorentzian density of states for each of the reservoirs
\beq
\label{lorentzian_dos}
\mathfrak{D}_r(\epsilon)=\frac{1}{2\pi}\frac{\kappa_r}{(\epsilon-\delta_r)^2+(\kappa_r/2)^2}.
\eeq
 This effective density of states describes the broadening of the normal modes playing the role of the reservoirs caused by their  coupling to the electromagnetic bath through the laser-cooling process, and also appears in the theory of spontaneous emission inside a leaky cavity~\cite{lorentzian_cavity}.

We have finally obtained and expression of  the transition rates, which can be compared to the conventional  orthodox theory of electronic quantum transport~\cite{single_electron_review}, and exploited to reavisit the discussion about the transport window at the beginning of this section. The rate   $\Gamma_{r{\rm M}}(\ell,\ell')$, describing the transfer of an energy quantum from the magnet onto the reservoirs, displays a bosonic amplification $1+\frak{n}_r(\epsilon)$, which differs crucially from the fermionic suppression $1-\frak{f}_r(\epsilon)$ of the  electronic case. This difference forbids the definition of an energy transport window similar to the electronic transport window of Fig.~\ref{fig_transport_window}{\bf (a)}, since the energy transfer does not require an empty level in the drain reservoir. The other crucial difference with respect to the orthodox theory of electronic transport, which assumes a featureless density of states of the metallic leads, is the appearance of a Lorenzian-shaped density of states $\mathfrak{D}_r(\epsilon)$ for the thermal reservoirs, whose centre and width can be controlled by tuning the parameters of the laser beams inducing the energy exchange~\eqref{energy_exchange}, and the laser cooling~\eqref{cooling_dissipator}, respectively. As depicted in Fig.~\ref{fig_transport_window}{\bf (b)}, one can envisage exploiting such density of states in order to engineer a similar transport window  for the flow of energy quanta through the quantum magnet. This opens a vast amount of possibilities of observing interesting  quantum effects in the transport of energy that had been  restricted to  electronic currents so far.

In the context of trapped ions, this transport scheme adds onto the toolbox of {\it sympathetic dissipative gadgets}, where the ability to control the effective density of states, or equivalently the spectral density, has been exploited to propose schemes for dissipation-assisted two-qubit gates~\cite{dissipation_assisted}, and for the dissipative generation of multi-particle entangled states~\cite{cooling_chain}.

\section{\bf Ising blockade  for energy transport}
\label{sec:blockade}

To illustrate the prospects of the trapped-ion energy transport scheme,  we apply the general formalism presented in Sec.~\ref{sec:trapped_ions_energy_tranport} to a phenomenon of particular relevance already discussed in the introduction of this manuscript: controlling energy transport at the single  quantum level by means of an analogue of the well-known Coulomb-blockade effect~\cite{coulomb_blockade,single_electron_review}. We discuss the energy transport through an Ising dimer connected to a pair of temperature-biased transport reservoirs. Moreover, its implementation with trapped-ion setups would be the cleanest possible experiment of the  transport toolbox.

%%%%%%%%%%%%%%%%%%%%%%%%%%%%%%%%%%%%%%%%%%%%%%%%%%%%%%%%%%%%
\begin{figure}
\centering
\includegraphics[width=0.9\columnwidth]{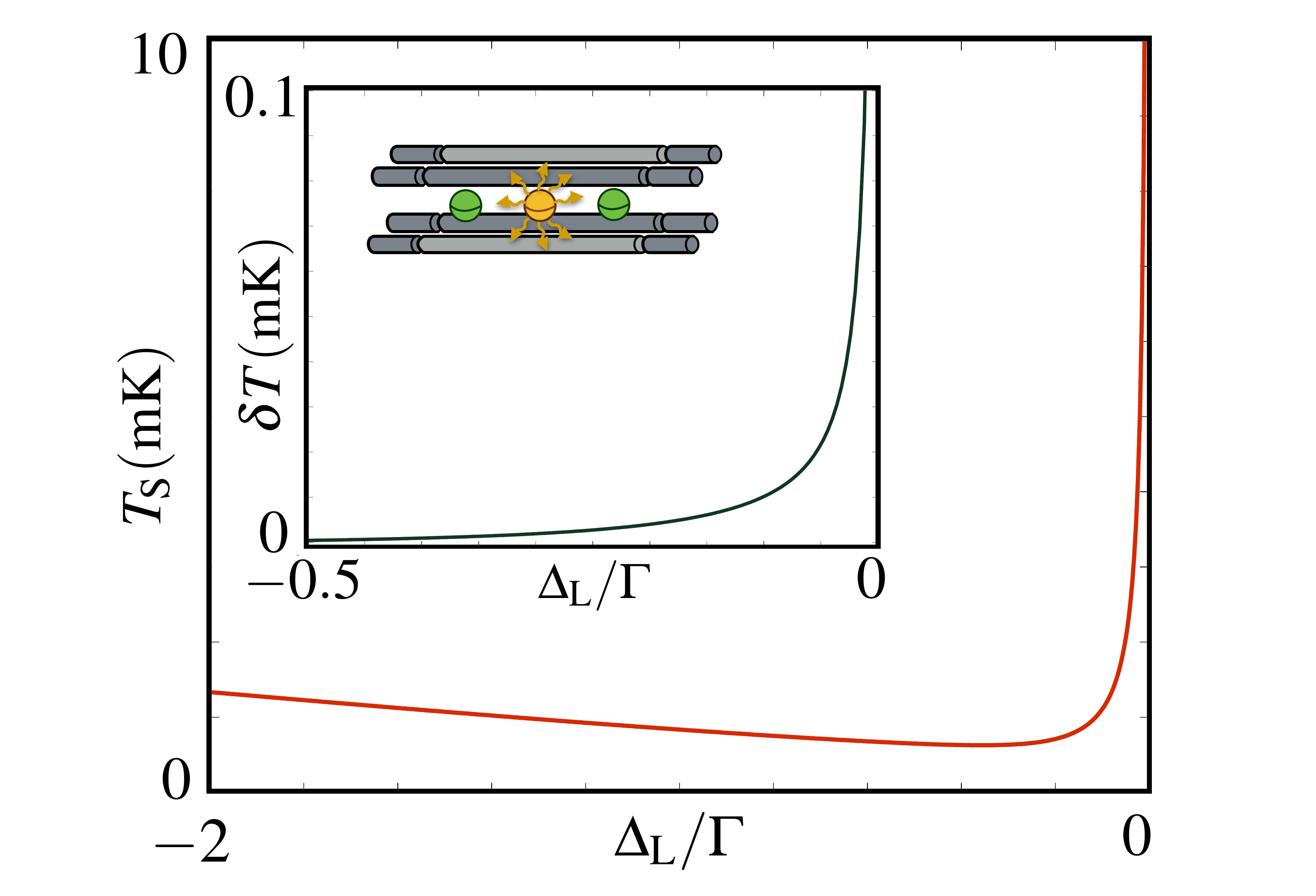}
\caption{ {\bf Temperatures of the laser-cooled effective reservoirs: } ({\bf main panel}) Temperature of the laser-cooled center-of-mass mode acting as the source reservoir, and temperature difference between the center-of-mass and egyptian mode ({\bf inset}) of a three-ion $^{25}{\rm Mg}^+$-$^{26}{\rm Mg}^+$-$^{25}{\rm Mg}^+$ crystal, as a function of the detuning of the laser with respect to a dipole allowed transition  of $^{26}{\rm Mg}^+$. Parameters considered: trap frequency $\omega_z/2\pi=1{\rm MHz}$, natural decay rate $\Gamma/2\pi=41.4{\rm MHz}$, cooling laser in a travelling-wave configuration with Rabi frequency $\Omega_{\rm L}=\Gamma/2$, and various detunings $\Delta_{\rm L}$ with respect to the $3S_{1/2}$-$3P_{1/2}$ transition.}
\label{fig_Ts}
\end{figure}
%%%%%%%%%%%%%%%%%%%%%%%%%%%%%%%%%%%%%%%%%%%%%%%%%%%%%%%%%%%%

For the sake of concreteness, let us consider a   three-ion   $^{25}{\rm Mg}^+$-$^{26}{\rm Mg}^+$-$^{25}{\rm Mg}^+$ crystal, where the inner ion is Doppler cooled by a travelling-wave  laser beam with a frequency  close to $^{26}{\rm Mg}^+$ dipole-allowed transition $3S_{1/2}$-$3P_{1/2}$. By the sympathetic cooling described in Sec.~\ref{sec:reservoirs}, the collective longitudinal modes are continuously pumped into the thermal state~\eqref{cooling_steady_state}. For our particular mixed-species crystal, these normal modes are only slightly perturbed with respect to those of a single-species crystal~\cite{normal_modes}. In particular, the  lowest- and highest-frequency modes, the so-called center-of-mass $n_{\rm S}=1$ and egyptian $n_{\rm D}=3$ modes,  are separated in frequencies by a large energy gap. Thus,  we can  address them individually with the red-sideband laser excitations described in Eq.~\eqref{energy_exchange}, such that these sympathetically-cooled modes play the role of the source-drain reservoirs of Eq.~\eqref{cooling_steady_state}, whose temperature bias is depicted in Fig.~\ref{fig_Ts}.

To apply the general formalism of  Sec.~\ref{sec:trapped_ions_energy_tranport}, let us start by diagonalising the Ising model~\eqref{Ising} for a vanishing transverse field, which describes the antiferromagnetic spin-spin interactions between the two outer ions  $^{25}{\rm Mg}^+$-$^{26}{\rm Mg}^+$- $^{25}{\rm Mg}^+$ mediated by the radial phonons. The eigenstates and energies are
\beq
\begin{split}
&\epsilon_{\ell\phantom{'}}=0\phantom{J},\hspace{2ex}\ket{\epsilon_{{\uparrow\downarrow}}}=\ket{{\uparrow\downarrow}}, \hspace{0.25ex} \ket{\epsilon_{{\downarrow\uparrow}}}=\ket{{\downarrow\uparrow}},\\
&\epsilon_{\ell'}=2J,\hspace{2ex}\ket{\epsilon_{{\uparrow\uparrow}}}=\ket{{\uparrow\uparrow}}, \hspace{0.25ex} \ket{\epsilon_{{\downarrow\downarrow}}}=\ket{{\downarrow\downarrow}}.
\end{split}
\eeq
To simplify the analysis  further, we exploit the  configuration of the isotopes, which leads to  symmetric displacements $\mathcal{M}^z_{1,n_r}=\mathcal{M}^z_{3,n_r}$ of the outer $^{25}{\rm Mg}^+$ ions in both normal modes.   If the wavelengths of the Raman beams that yield  the red-sideband excitations~\eqref{energy_exchange} are adjusted adequately with respect to the inter-ion distance, one finds that the energy-exchange couplings in that expression are homogeneous  $g_{i,r}=g_r$. In this situation, the singlet state $\ket{\epsilon_{\rm S}}=(\ket{{\uparrow\downarrow}}-\ket{{\downarrow\uparrow}})/\sqrt{2}$ gets decoupled from the transport channels. This is the non-equilibrium transport analogue of the sub-radiant decay channel of two distant atoms coupled to the electromagnetic bath, and the associated singlet dark state in the spontaneous emission~\cite{superadiance}. 

%%%%%%%%%%%%%%%%%%%%%%%%%%%%%%%%%%%%%%%%%%%%%%%%%%%%%%%%%%%%

\begin{figure}
\centering
\includegraphics[width=0.8\columnwidth]{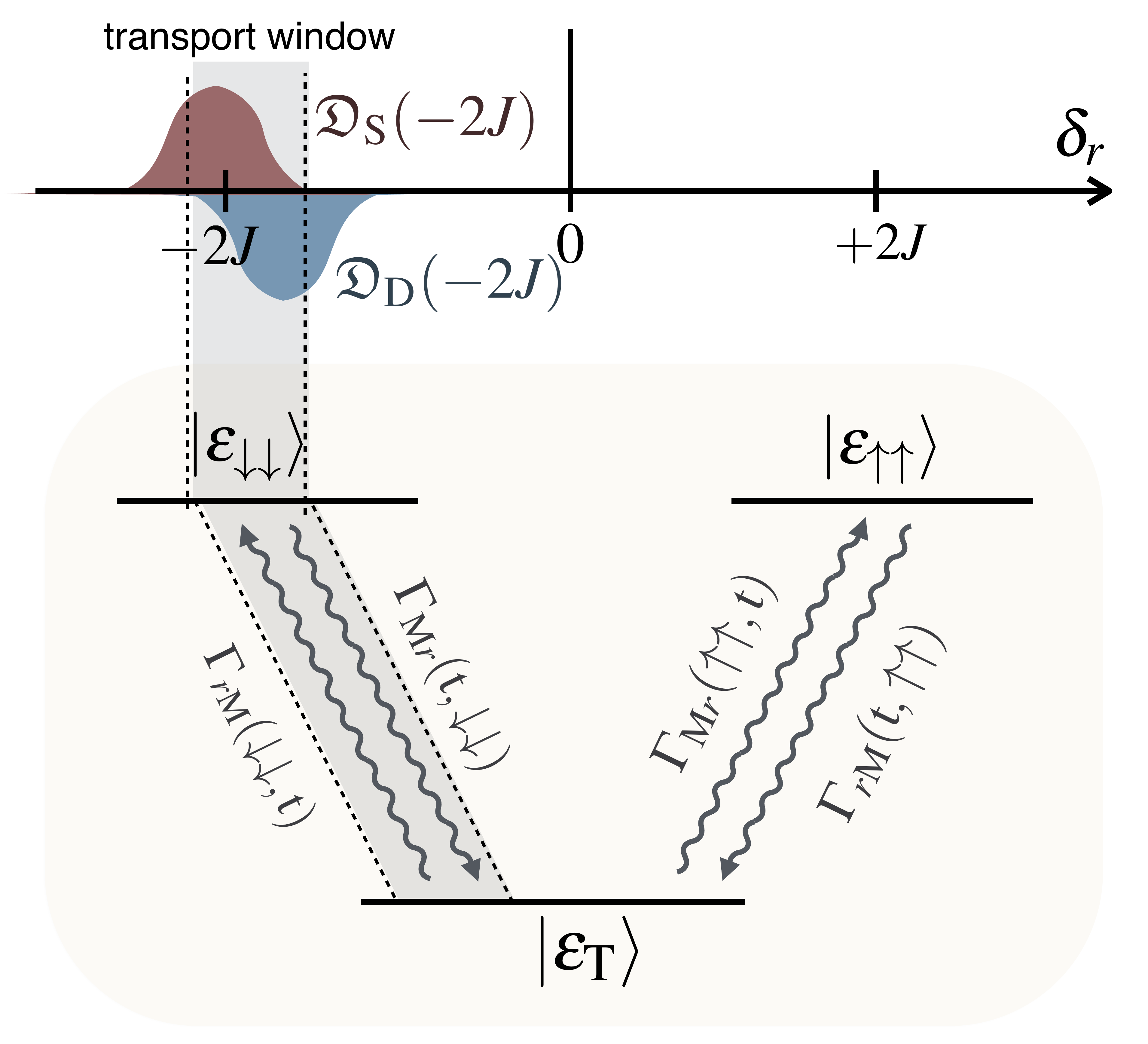}
\caption{ {\bf Energy transport through an Ising dimer:} The energy transport through a couple of interacting Ising spins can occur via two possible transport channels that involve the transitions $\ket{\epsilon_{\rm T}}\leftrightarrow\ket{\epsilon_{\uparrow\uparrow}}$, or $\ket{\epsilon_{\rm T}}\leftrightarrow\ket{\epsilon_{\downarrow\downarrow}}$ in the Ising magnet, induced by the absorption/emission of an energy quantum from/into the biased reservoirs. Controlling the Lorenztian density of states of the synthetic reservoirs,  a single transport channel can be selected (e.g. $\ket{\epsilon_{\rm T}}\leftrightarrow\ket{\epsilon_{\downarrow\downarrow}}$, when the Lorentzians are centred  at negative frequencies by using negative laser detunings).     }
\label{fig_ising_scheme}
\end{figure}
%%%%%%%%%%%%%%%%%%%%%%%%%%%%%%%%%%%%%%%%%%%%%%%%%%%%%%%%%%%%
Accordingly, the energy transport gets simplified, as we only need to consider three states
\beq
\begin{split}
&\epsilon_{\ell\phantom{'}}=0\phantom{J},\hspace{2ex}\ket{\epsilon_{\rm T}}=(\ket{{\uparrow\downarrow}}+\ket{{\downarrow\uparrow}})/\sqrt{2},\\
&\epsilon_{\ell'}=2J,\hspace{2ex}\ket{\epsilon_{\uparrow\uparrow}}=\ket{{\uparrow\uparrow}}, \hspace{0.25ex} \ket{\epsilon_{\downarrow\downarrow}}=\ket{{\downarrow\downarrow}}.
\end{split}
\eeq
Due to the particular form of the energy-exchange term~\eqref{energy_exchange}, the only transitions allowed form the ${\bf V}$-scheme depicted in Fig.~\ref{fig_ising_scheme}, which corresponds to a couple of transport channels for the flow of energy through the Ising dimer. Using the general expressions~\eqref{rates}, we find that the particular rates connecting  $\ket{\epsilon_{\rm T}}\leftrightarrow\ket{\epsilon_{\uparrow\uparrow}}$ are governed by the effective density of states evaluated at positive  frequencies
\beq
\label{channel_1}
\begin{split}
\Gamma_{r{\rm M}}({\rm T},\uparrow\uparrow)&=4\pi|g_r|^2\frak{D}_r(+2J)(1+\frak{n}_r(\omega_r)),\\
\Gamma_{{\rm M}r}({\uparrow\uparrow},{\rm T})&=4\pi|g_r|^2\frak{D}_r(+2J)\frak{n}_r(\omega_r).
\end{split}
\eeq
 Conversely, the rates connecting    $\ket{\epsilon_{\rm T}}\leftrightarrow\ket{\epsilon_{\downarrow\downarrow}}$  depend on the effective density of states evaluated at negative  frequencies
 \beq
 \label{channel_2}
\begin{split}
\Gamma_{r{\rm M}}(\downarrow\downarrow,{\rm T})&=4\pi|g_r|^2\frak{D}_r(-2J)(1+\frak{n}_r(\omega_r)),\\
\Gamma_{{\rm M}r}({\rm T},{\downarrow\downarrow})&=4\pi|g_r|^2\frak{D}_r(-2J)\frak{n}_r(\omega_r).
\end{split}
\eeq

Provided that {\it (i)} the centre of the Lorentzian densities~\eqref{lorentzian_dos}, which are controlled by the red-sideband detunings $\delta_r$~\eqref{laser_detunings}, lie at negative frequencies $\delta _{r}\approx -2J$, and {\it (ii)} the width of the Lorentzian densities~\eqref{lorentzian_dos}, which are controlled by the laser-cooling rate~\eqref{cond_reservoir} given by the rates~\eqref{heating_cooling},   fulfil $\kappa_r\ll J$, then one can ensure that only the transport channel $\ket{\epsilon_{\rm T}}\leftrightarrow\ket{\epsilon_{\downarrow\downarrow}}$  will be active [cf. Fig.~\ref{fig_ising_scheme}]. In this regime, the transport master equation~\eqref{transport_meq} for the single-channel populations can be easily solved, leading to the steady state $\rho_{\rm m}^{\rm eq}=\sum_{\ell,\ell'}\rho^{\rm eq}_{\ell,\ell'}\ket{\ell}\bra{\ell'}$ that displays the following diagonal terms
\beq
\label{eq_pop}
\begin{split}
\rho_{\downarrow\downarrow,\downarrow\downarrow}^{\rm eq}=\frac{\Gamma_{{\rm SM}}(\downarrow\downarrow,{\rm T})+\Gamma_{{\rm DM}}(\downarrow\downarrow,{\rm T})}{\Gamma_{\rm tot}},\\
\rho_{\rm T,T}^{\rm eq}=\frac{\Gamma_{{\rm MS}}({\rm T},\downarrow\downarrow)+\Gamma_{{\rm MD}}({\rm T},\downarrow\downarrow)}{\Gamma_{\rm tot}},
\end{split}
\eeq
where we  introduced $\Gamma_{\rm tot}=\sum_r\Gamma_{{\rm M}r}({\rm T},\downarrow\downarrow)+\sum_r\Gamma_{r{\rm M}}(\downarrow\downarrow,{\rm T})$, and assumed that the initial state of the quantum magnet has no population on the transport dark state (e.g. $\rho_{\rm m}(0)=\ket{\downarrow\downarrow}\bra{\downarrow\downarrow}$). Moreover, calculating the current of energy quanta flowing through the synthetic quantum magnet from the source reservoir,
$
I_{\rm S}=\Gamma_{{\rm MS}}({\rm T},\downarrow\downarrow)\rho_{\downarrow\downarrow,\downarrow\downarrow}^{\rm eq}-\Gamma_{{\rm SM}}(\downarrow\downarrow,{\rm T})\rho_{\rm t,t}^{\rm eq}
$,
leads to the following expression
\beq
\label{current_app}
I_{\rm S}=\frac{4\pi|g_{\rm S}g_{\rm D}|^2\frak{D}_{\rm S}(-2J)\frak{D}_{\rm D}(-2J)}{\sum_r|g_{r}|^2\frak{D}_{r}(-2J)(1+2\frak{n}_{r}(\omega_{r}))}\big(\frak{n}_{\rm S}\big(\omega_{\rm S})-\frak{n}_{\rm D}(\omega_{\rm D})\big).
\eeq
Let us highlight the following physical predictions of this expression: {\it (i)} energy will flow through the synthetic magnet from the hotter source reservoir into the colder drain reservoir, since $\frak{n}_{\rm S}(\omega_{\rm S})>\frak{n}_{\rm D}(\omega_{\rm D})$. {\it (ii)} In general, the  energy current violates Fourier's law of heat conduction since the proportionality coefficient of  $I_{\rm S}\propto\big(\frak{n}_{\rm S}\big(\omega_{\rm S})-\frak{n}_{\rm D}(\omega_{\rm D})\big)$ does also depend on the temperature of the reservoirs, and not only on features of the system. Note that violations of Fourier's law abound at the microscopic level~\cite{fourier_law}, similarly to the violations of Ohm's law in the electronic transport though nano-devices~\cite{nazarov_blanter}. {\it (iii)} The energy current depends on the overlap of the Lorentzian densities of states of both reservoirs  $I_{\rm S}\propto\frak{D}_{\rm S}(-2J)\frak{D}_{\rm D}(-2J)$ evaluated at an energy that depends on the Ising interaction strength. Unless the transport window, defined by the overlap of both  Lorentzians, contains the region around $-2J$, energy transport will be blockaded $I_{\rm S}\approx0$ since the spins $\ket{\epsilon_{\downarrow\downarrow}}$ do not have the required energy to overcome the Ising gap to populate the Bell state $\ket{\epsilon_{\rm t}}$. This phenomenon is the energy-transport analogue of Coulomb blockade in electronic devices, and it can be understood as an Ising blockade mechanism of energy transport through the synthetic quantum magnet.

%%%%%%%%%%%%%%%%%%%%%%%%%%%%%%%%%%%%%%%%%%%%%%%%%%%%%%%%%%%%

\begin{figure}
\centering
\includegraphics[width=0.8\columnwidth]{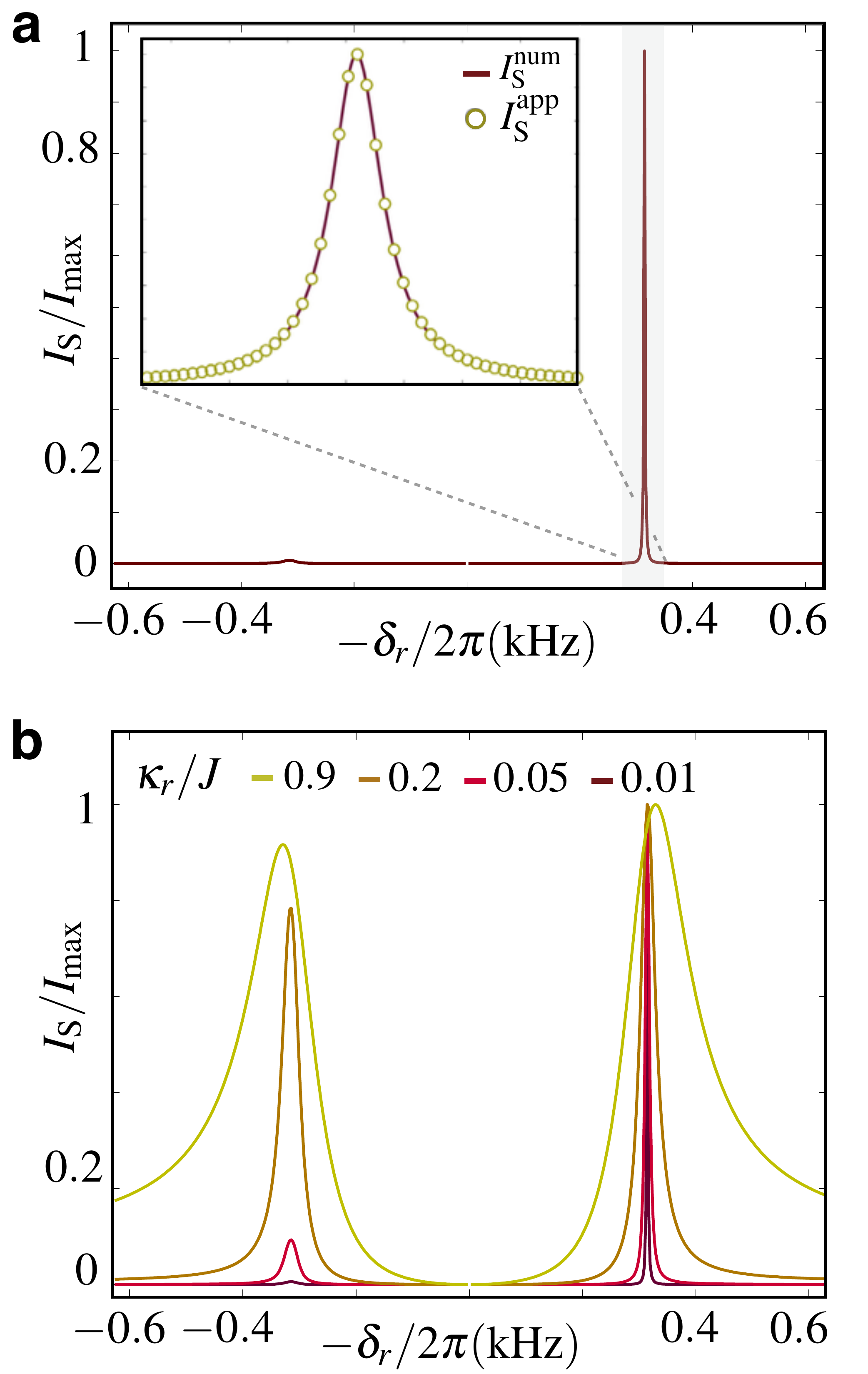}
\caption{ {\bf Ising blockade of energy transport :} Asymptotic energy current through the quantum magnet as a function of the detunings of the energy-exchange mechanism. We consider the regime of a well-resolved single transport channel $\kappa_r\ll J$ in {\bf (a)}, and study how the current is changed as one leaves this regime $\kappa_r\to J$ in {\bf (b)}. Parameters considered:  trap frequency $\omega_z/2\pi=1{\rm MHz}$, natural decay rate $\Gamma/2\pi=41.4{\rm MHz}$, cooling laser in a travelling-wave configuration with detuning $\Delta_{\rm L}=\Gamma_\tau/2$, and various cooling strengths $\Omega_{\rm L}/\gamma_\tau\in\{4\cdot 10^{-3}, 8\cdot 10^{-3},1.6\cdot 10^{-2},3.2\cdot 10^{-2}\}$, which lead to the cooling rates $\kappa_r$ of   {\bf (b)}.   For the synthetic magnet, we consider far-detuned state-dependent dipole force with detuning $\delta_{\rm L}=0.1\omega_x$, for a radial trap frequency $\omega_x/2\pi=4{\rm MHz}$,   and strength $Fx_0=0.1\delta_{\rm L}$, leading to antiferromagnetic Ising couplings $J/2\pi=0.16
${\rm kHz}.  }
\label{fig_ising_blockade}
\end{figure}
%%%%%%%%%%%%%%%%%%%%%%%%%%%%%%%%%%%%%%%%%%%%%%%%%%%%%%%%%%%%

We now test the validity of Eq.~\eqref{current_app} by a numerical comparison with the full transport master equation~\eqref{transport_meq} containing the two transport channels with rates~\eqref{channel_1}-\eqref{channel_2} for the $^{25}{\rm Mg}^+$-$^{26}{\rm Mg}^+$-$^{25}{\rm Mg}^+$ crystal. In Fig.~\ref{fig_ising_blockade}{\bf (a)}, we represent in a solid line the calculated asymptotic current after solving numerically the master equation as a function of the inverted  laser detunings $-\delta_r$ of the red-sideband excitations, considering  an initial state $\rho_{\rm m}(0)=\ket{\downarrow\downarrow}\bra{\downarrow\downarrow}$. We observe that the energy transport is Ising-blockaded, except for  detunings $\delta_{r}\approx -2J$, where the transport window  contains  the channel~\eqref{channel_2}, and energy is allowed to flow through the synthetic magnet.  In the inset of this figure, we compare the Ising-blockaded oscillation with the theoretical prediction~\eqref{current_app} based on the single-channel approximation, and we observe a very good agreement. As occurs in an electronic single-electron transistor~\cite{single_electron_review}, where electrons tunnel sequentially   through a metallic island one by one when an applied gate voltage sets the Coulomb-blockaded channel within the transport window, the energy quanta also tunnel one by one through our synthetic  magnet when the laser detunings fulfil $\delta_r\approx -2J$. Since these detunings depend on the transition frequency  of the electronic levels conforming the spins $\omega_0$, and this frequency is shifted through the Zeeman effect,  the role of the gate voltage in the single-electron transistor can be played by an   external magnetic field in our setup, leading to an analogue single-quantum energy transistor.

In Fig.~\ref{fig_ising_blockade}{\bf (b)}, we study how the Ising blockade is lifted as the transport window becomes wider and wider. We observe that, as one increases the laser-cooling rates corresponding to the Lorentzian widths $\kappa_r$, the Ising blockade peak broadens signalling that energy transport can also occur within a larger bandwidth. Moreover, the second transport channel gets activated as $\kappa_r\to J$, and we observe how a second peak in the energy current rises at negative detunings.

\section{\bf Energy current measurements}
\label{sec:measurement}

The experimental study of energy  transport  through nano-structures is hampered by the lack of a controllable device that can measure   energy/heat currents directly~\cite{rmp_heat_transport}. This question was addressed in Ref.~\cite{heat_transport_sympathetic_toolbox_ions}, which considered a Ramsey scheme  whereby an additional qubit serves as a quantum sensor to probe the relevant observables. The qubit gathers information  about the mean energy  current, and its fluctuation  spectrum at zero frequency, while minimally perturbing the non-equilibrium steady state that supports the energy current. Unfortunately, this scheme is quite specific to the situation where energy flows via  the vibrational excitations of the ion crystal. Moreover, the measurement scheme requires additional laser beams to implement a spin-dependent version of the photon-assisted tunnelling of vibrational excitations in the ion crystal~\cite{pat_ions}, which is essential  to map the information of the energy current onto the phase  of the qubit.

In this section, we take a different approach, and devise a more direct measurement scheme for the energy current at the expense of destroying the non-equilibrium steady-state after each measurement. Therefore,  the steady state of the biased quantum magnet must be prepared before each of the measurements.  The main idea is that if  the coupling of the synthetic magnet to the drain reservoir~\eqref{energy_exchange} is suddenly switched off,  the evolution of the magnet populations at very short times encodes the expectation value of the energy current.

In order to prove the above statement, let us consider the  Ising-blockaded magnet of the previous section, which is described by the transport master equation~\eqref{transport_meq} with the rates of the relevant  channel~\eqref{channel_2}. After the magnet has equilibrated to the state~\eqref{eq_pop} during the interval $0<t<t_{\rm q}$, such that $t_{\rm q} \gg t_{\rm eq}$ and $t_{\rm eq}$ is the equilibration time, we quench the system by switching off its coupling to the drain reservoir at $t=t_{\rm q}$, namely $g_{\rm D}(t_{\rm q})=0$. The evolution of the populations for $t> t_{\rm q}$ is described by the following rate equations
\beq
\begin{split}
\frac{{\rm d} \rho_{\downarrow\downarrow,{\downarrow\downarrow}}(t)}{{\rm d} t}=\Gamma_{\rm SM}(\downarrow\downarrow,{\rm T}) \rho_{\rm T,T}(t)-\Gamma_{\rm MS}({\rm T},\downarrow\downarrow) \rho_{\downarrow\downarrow,{\downarrow\downarrow}}(t),\\
\frac{{\rm d} \rho_{\rm T,T}(t)}{{\rm d} t}=\Gamma_{\rm MS}({\rm T},\downarrow\downarrow) \rho_{\downarrow\downarrow,{\downarrow\downarrow}}(t)-\Gamma_{\rm SM}(\downarrow\downarrow,{\rm T}) \rho_{\rm T,T}(t).
\end{split}
\eeq
By formally integrating these equations, and introducing the solution iteratively a couple of times, one finds that
\beq
\begin{split}
\rho_{\downarrow\downarrow,{\downarrow\downarrow}}(t_{\rm q}+\Delta t))&=\rho_{\downarrow\downarrow,{\downarrow\downarrow}}(t_{\rm q})\\
&+\Delta t \left(\Gamma_{{\rm SM}}(\downarrow\downarrow,{\rm T})\rho_{\rm t,t}(t_{\rm q})-\Gamma_{{\rm MS}}({\rm T},\downarrow\downarrow)\rho_{\downarrow\downarrow,\downarrow\downarrow}(t_{\rm q})\right)\\
&+\mathcal{O}(\Delta t^2).
\end{split}
\eeq
Since the  state $\rho_{\rm m}(t_{\rm q})$ at the beginning of the quench is the equilibrium state~\eqref{eq_pop}, we find that the time evolution at very short times fulfils
\beq
\label{en_current_obs}
I_{\rm S}=\frac{\rho_{\downarrow\downarrow,{\downarrow\downarrow}}(t_{\rm q})-\rho_{\downarrow\downarrow,{\downarrow\downarrow}}(t_{\rm q}+\Delta t))}{\Delta t}, \hspace{2ex}\Delta t\ll \Gamma_{{\rm SM}}^{-1},\Gamma_{{\rm MS}}^{-1},
\eeq
where $I_{\rm S}$ is precisely the expectation value of the energy current in Eq.~\eqref{current_app}. 

The measurement scheme for the energy current can be thus described as follows: {\it (i)} Prepare   the magnetic system in state $\rho_{\rm m}(0)=\ket{\downarrow\downarrow}\bra{\downarrow\downarrow}$ by optical pumping~\cite{wineland_review}, and let it equilibrate  with   the biased laser-cooled  modes playing the role of the reservoirs for a time $t_{\rm q}$. {\it (ii) } Switch off all the laser couplings, such that the populations of the quantum magnet get frozen at $t=t_{\rm q}$. {\it (iii)}  Measure the  spin population $\rho_{\downarrow\downarrow,\downarrow\downarrow}(t_{\rm q})=\bra{\downarrow\downarrow}\rho_{\rm m}(t_{\rm q})\ket{\downarrow\downarrow}$ through state-dependent fluorescence in a cycling transition of the ion crystal~\cite{wineland_review}. {\it (iv)} Repeat the above steps $\it (i)-(iii)$, but quenching the coupling to the drain reservoir precisely at $t=t_{\rm q}$, and letting the system evolve for a very short additional time $\Delta t\ll\Gamma_{\rm SM}^{-1},\Gamma_{\rm MS}^{-1}$ before switching off all laser couplings to freeze the magnetic populations, thus obtaining $\rho_{\downarrow\downarrow,\downarrow\downarrow}(t_{\rm q}+\Delta t)=\bra{\downarrow\downarrow}\rho_{\rm m}(t_{\rm q}+\Delta t)\ket{\downarrow\downarrow}$.   {\it (iv)} Infer the energy current from the ratio of Eq.~\eqref{en_current_obs}. 

According to this scheme, one can validate  our derivation of the energy current~\eqref{current_app}, and the quantum phenomenon of Ising blockade in the transport of energy. Let us note that it is also possible to measure how the system equilibrates with all the related transient phenomena, by sweeping over different values of $t_{\rm q}$. The available  time resolution in switching on/off the laser couplings, which can easily reach sub-$\mu$s, is more than enough in the context of the  slower transport dynamics of our scheme. We finally remark that, even if decoherence affects the internal degrees of freedom via external fluctuating magnetic fields, these fields are typically homogeneous along such small ion crystals, and will not modify the definition of the relevant transport channels. Moreover, since the dynamics of the populations in the transport is described by  rate equations, such that coherences between the different transport channels are not relevant, the predicted Coulomb blockade in the energy transport should be insensitive to such sources of decoherence. Only the presence of fluctuating magnetic-field gradients can induce a coupling between the triplet and singlet levels, such that a more general description of the transport channels might be required. Anyhow, since the Coulomb-blockade effect relies on the energetic splitting of the different levels, its signatures should still be present in this situation.

\section{\bf Conclusions and Outlook}
\label{sec:outlook}

We have proposed a scheme to study the quantum transport of energy through synthetic quantum magnets implemented with Coulomb crystals of trapped atomic ions. By exploiting sympathetic cooling in a mixed Coulomb crystal,  a pair of temperature-biased thermal reservoirs can be mimicked via two laser-cooled longitudinal modes of the crystal. The analogue of the microscopic scattering mechanism  that leads to the energy exchange between lattice vibrations and the magnetic moments required for energy transport in in solids can be realised by resolved laser-driven red-sideband couplings, whereas the  interactions within the magnet can be designed at will by phonon-mediated spin-spin couplings induced by far-detuned state-dependent dipole forces.

We have derived a general quantum master equation for the transport of energy, and applied it to the particular case of an Ising dimer, where we have found the energy-transport counterpart of the Coulomb blockade mechanism. This opens the possibility of exploring a variety of quantum effects in energy/heat transport that had been previously restricted to the realm of electronic transport. By quenching the coupling of the magnet to the drain reservoir, and  measuring certain populations by  state-dependent fluorescence of the ion crystal for very short times, we have shown that the energy current can be inferred, and the prediction of the blockade can be addressed in a  trapped-ion experiment.

The proposed toolbox can also address effects that go beyond the derived quantum master equation, and  would require a systematic extension to higher orders of  the adiabatic elimination in Eq.~\eqref{ad_elim}. For instance, since the ratio of the energy-exchange coupling and the Ising interaction strength can be modified at will, one can study the effects of energy co-tuneling in the Ising-blockaded regime or, ultimately, the possibility of observing the energy counterpart of the Kondo effect in quantum electronic transport~\cite{bruus_flensberg}. 

By adding a transverse field to the Ising model, and considering larger ion chains, one could explore the consequences of quantum phase transition in the energy transport, and even search for the predicted signatures of quantum criticality in the observed transverse currents~\cite{ising_chain_transverse_transport}. 

\ifcheckpagelimits
\else

{\bf \acknowledgements}
A. B. acknowledges support from Spanish MINECO Project FIS2012-33022, and CAM regional research consortium QUITEMAD+. T.S. is supported by DFG (SCHA 973). A. B. thanks the hospitality of FRIAS (Freiburg Institute of Advanced Science) within the research focus on ``Designed quantum transport in complex materials'', where parts of this work were developed.

\end{document}